\documentstyle{article}

\catcode`\@=11

\@addtoreset{equation}{section}
\catcode`\@=12
\def\sgn{\mathop{\rm sgn}\nolimits}

\font\bbb=msbm10
\def\C{\hbox{\bbb C}}
\def\R{\hbox{\bbb R}}
\newtheorem{Theorem}{Theorem}[section]
\newtheorem{Lemma}{Lemma}[section]

\newtheorem{Corollary}{Corollary}[section]
\begin{document}

\title{B\"{a}cklund and Darboux transformations for the nonstationary
Schr\"{o}dinger equation\thanks{Work supported in part by PRIN 97
``Sintesi'' and grant of RFBR \#96-01-00344.}}
\author{M. Boiti, F. Pempinelli, A. Pogrebkov
\thanks{Permanent address: Steklov Mathematical Institute, Gubkin str., 8,
Moscow, 117966, GSP-1, Russia; e-mail pogreb@mi.ras.ru}, and B.Prinari \\
{}Dipartimento di Fisica dell'Universit\`{a} \\
Sezione INFN, 73100 Lecce, ITALY\thanks{e-mail boiti@le.infn.it}}
\maketitle

\begin{abstract}
Potentials of the nonstationary Schr\"{o}dinger operator constructed by
means of $n$ recursive B\"{a}cklund transformations are studied in detail.
Corresponding Darboux transformations of the Jost solutions are introduced.
We show that these solutions obey modified integral equations and present
their analyticity properties. Generated transformations of the spectral data
are derived.
\end{abstract}

\section{Introduction}

In this article we continue (see~\cite{nsreshi}--\cite{towards}) our
investigation into the direct and inverse scattering transform for the
nonstationary Schr\"{o}dinger operator
\begin{equation}
L=i\partial _{x_{2}}^{}+\partial _{x_{1}}^{2}-u(x_{1}^{},x_{2}^{}),
\label{1}
\end{equation}
in the case in which $u$ is a real function with ``ray'' type behavior. More
exactly, $u$ is supposed to be rapidly decaying in all directions on the
$x$-plane with the exception of some finite number of directions, where it has
finite and nontrivial limits, i.e.
\begin{equation}
u_{n,\pm }^{}(x_{1}^{})=\lim_{x_{2}\rightarrow \pm \infty}
u(x_{1}^{}-2\mu _{n}^{}x_{2},x_{2}^{}),\qquad n=1,2,\dots ,N,
\label{u+-}
\end{equation}
for $N$ real constants $\mu _{n}$. Spectral theory of operator $L$ with
potential of this class is interesting {\it per se} and because it is
associated~\cite{dryuma74,zakharov74} to the Kadomtsev--Petviashvili
equation~\cite{Kadomtsev-P}---in its version called KPI---
\begin{equation}
(u_{t}^{}-6uu_{x_{1}}^{}+u_{x_{1}x_{1}x_{1}}^{})_{x_{1}}^{}=
3u_{x_{2}x_{2}}^{}.
\label{KPI}
\end{equation}

The mentioned extension of the spectral theory of the nonstationary
Schr\"{o}dinger operator would provide possibility to extend
correspondingly the class of solutions of the KPI equation, including
potentials with asymptotic one dimensional behavior. On the other side such
spectral theory is essential for investigation by the inverse scattering
transform method of strongly localized soliton solutions and their
interaction with background for the Davey--Stewartson I equation, where
operator~(\ref{1}) controls behavior of the ``boundaries at infinity'' of an
auxiliary function for this equation (see~\cite{localized}--\cite{fractal}
for details).

Spectral theory of the operator~(\ref{1}) with ray potential is essentially
more involved than the standard case of rapidly decaying potential.
Indeed, in the standard case~\cite{ZManakov}--\cite{KPJMP} one defines the
Jost solution $\Phi (x,k)$ as the solution of the nonstationary
Schr\"{o}dinger equation,
\begin{equation}
(i\partial _{x_{2}}^{}+\partial _{x_{1}}^{2}-u(x))\phi (x,k)=0,
\label{Schru}
\end{equation}
that is analytic in the complex plane of the spectral parameter $k$, $k_{\Im
}\neq 0$, and normalized at infinity by the condition that for the function
\begin{equation}
\chi (x,k)=e_{}^{ikx_{1}+ik^{2}x_{2}}\Phi (x,k)  \label{3}
\end{equation}
we have
\begin{equation}
\lim_{k\rightarrow \infty ,\,k_{\Im }\neq 0}\chi (x,k)=1.  \label{asymptk}
\end{equation}
This function can be given as the solution of the integral equation~\cite
{Manakov}
\begin{equation}
\chi (x,k)=1+\int dx'G_{0}^{}(x-x',k)u(x')\chi
(x',k),  \label{4}
\end{equation}
where $G_{0}(x,k)$ is the Green's function
\begin{equation}
G_{0}^{}(x,k)=\frac{\mathop{\rm sgn}\nolimits x_{2}^{}}{2\pi i}\int
d\alpha \theta (\alpha k_{\Im }^{}x_{2}^{})e_{}^{i\alpha x_{1}-i\alpha
(\alpha -2k)x_{2}},\qquad k\in \C,  \label{6}
\end{equation}
of the differential equation~(\ref{1}) with zero potential,
\begin{equation}
(i\partial _{x_{2}}^{}+\partial _{x_{1}}^{2}-2ik\partial
_{x_{1}}^{})G_{0}^{}(x,k)=\delta (x).  \label{61}
\end{equation}
Solvability of these differential equations under some small norm
assumptions was proved in~\cite{Xin} and thanks to~(\ref{4}) it is easy to
show that $\chi (x,k)$ has the asymptotic behavior
\begin{equation}
\lim_{\left| x_{1}\right| \rightarrow \infty }\chi (x,k)=1  \label{asymptx}
\end{equation}
on the $x$-plane, $k_{\Im }\neq 0$, independently on the direction. The Jost
solution $\Phi (x,k)$ defined by~(\ref{3}) obeys~\cite{kpshort,KPJMP}
the following normalization and completeness conditions
\begin{eqnarray}
&&\int dx_{1}^{}\overline{\Phi (x_{1}^{},x_{2}^{},k+p)}\Phi
(x_{1}^{},x_{2}^{},\bar{k})=2\pi \delta (p),\qquad p\in \R,
\label{81} \\
&&\int dk_{\Re }^{}\overline{\Phi (x_{1}',x_{2}^{},k)}\Phi
(x_{1}^{},x_{2}^{},\bar{k})=2\pi \delta (x_{1}^{}-x_{1}').
\label{82}
\end{eqnarray}

The Jost solution is an analytic function of $k$ in the complex plane,
$k_{\Im}\neq 0$. It has finite limits on the real axis,
\begin{equation}
\Phi _{}^{\pm }(x,k)=\lim_{k_{\Im }\rightarrow \pm 0}\Phi(x,k).  \label{sd1}
\end{equation}
By~(\ref{81}) these boundary values obey the normalization conditions
\begin{equation}
\int dx_{1}^{}\overline{\Phi _{}^{\pm }(x_{1}^{},x_{2}^{},k)}\Phi
_{}^{\mp }(x_{1}^{},x_{2}^{},p)=2\pi \delta (p-k),\qquad p,k\in \R.
\label{sd2}
\end{equation}
Spectral data are introduced as measure of the departure from
analyticity of the Jost solutions. In literature there exists a variety of
definitions of spectral data. In what follows we use the one suggested
in~\cite{KP,kpshort,KPJMP}:
\begin{equation}
{\cal F}(k,p)=\frac{1}{2\pi }\int dx_{1}'\overline{\Phi _{}^{+}(x_{1}',
x_{2}^{},k)}\Phi _{}^{+}(x'_{1},x_{2}^{},p)-\delta (p-k),
\qquad p,k\in \R.  \label{sd3}
\end{equation}
It is easy to check that ${\cal F}(k,p)$ is independent on $x_{2}$.
Condition of reality of the potential $u$ in~(\ref{1}) is equivalent to the
self-adjointness of the integral operator with kernel ${\cal F}(k,p)$.
Now by~(\ref{82}) we have
\begin{equation}
\Phi _{}^{+}(x,k)=\Phi _{}^{-}(x,k)+\int dp\Phi _{}^{-}(x,p){\cal F}(p,k).
\label{sd4}
\end{equation}
Thus the inverse scattering transform is the nonlocal Riemann--Hilbert
problem of construction of the function $\Phi (x,k)$ analytic in the upper
and bottom half planes with normalization~(\ref{3}) and~(\ref{asymptk})
and discontinuity at the real axis given by~(\ref{sd4}).

It was mentioned in~\cite{total,proc95,KPlett,towards} that the integral
equation~(\ref{4}) cannot be applied in the case of a potential $u(x)$ not
vanishing in all directions at large distances as the Green's function is
slow decaying at space infinity. In~\cite{towards} the following
modification of this integral equation was suggested:
\begin{equation}
\chi (x,k)=1+ \hspace{-2pt} \int\limits_{-k_{\Im }\infty }^{\qquad x_{1}}
\hspace{-6pt} dy_{1}^{}\int dx'\partial_{y_{1}}G_{0}^{}(y_{1}^{}-x_{1}',
x_{2}^{}-x'_{2},k)u(x')\chi (x',k),
\label{12}
\end{equation}
where the order of operations is explicitly prescribed. Here and
below we use notations of the type $k_{\Im }\infty $ in the limits of
integrals to indicate the sign of infinity. If the solution of this equation
exists and is bounded, then like in the one dimensional case
\begin{equation}
\lim_{x_{1}\rightarrow -k_{\Im }\infty }\chi (x,k)=1,\qquad k_{\Im }\neq 0,
\label{asx}
\end{equation}
while in contrast to~(\ref{asymptx}) it can be different from 1 in the
opposite direction. This modified integral equation is applicable to the
simplest case of a potential of type~(\ref{u+-}), i.e.\ to the case
$u(x)=u(x_{1})$ and it is trivial to check that it gives the standard
(see~\cite{NMPZ}) one dimensional equation for the Jost solution.
Nevertheless, the full description of the solutions of the Eq.~(\ref{12})
with potentials of the class~(\ref{u+-}) is absent till now. Only multiple
pure soliton solutions were constructed
in~\cite{KPsolitonsfirst,generalKPsolitons}. In~\cite{FokasPogreb}
and~\cite{towards} it was shown that solutions of this equation can have
additional cuts in the complex plane of the spectral parameter. Because of
this we are studying here the special but rather wide subclass of potentials
of  type~(\ref{u+-}) that is obtained by applying recursively the so called
binary B\"{a}cklund transformations~\cite{matveev} with complex spectral
parameter to a decaying potential. As we are interested in the spectral
characteristics of potentials $u$ having nontrivial limits, we study also the
corresponding Darboux transformations furnishing the Jost solutions of the
transformed potentials and their analytical properties as well as
transformations of the spectral data.

We have at our disposal in~\cite{matveev} and~\cite{salle} a rather simple
and transparent method for performing binary B\"{a}cklund transformations of
the potential $u$ and corresponding Darboux transformations of solutions
of~(\ref{1}). Let $\phi (x,k)$ be a solution of the nonstationary
Schr\"{o}dinger equation~(\ref{Schru}) with potential $u(x)$. Then the
transformed potential is equal to
\begin{equation}
\tilde{u}(x)=u(x)-2\partial _{x_{1}}^{2}\log \Delta (x),  \label{53}
\end{equation}
where
\begin{equation}
\Delta (x)=\int\limits^{x_{1}}dx_{1}'|\phi (x'_{1},x_{2}^{},\lambda )|
_{}^{2}.  \label{a55}
\end{equation}
The Darboux transform of $\phi(x,k)$,
\begin{equation}
\tilde{\phi}(x,k)=\phi (x,k)-\frac{\phi (x,\lambda )}{\Delta (x)}
\int\limits^{x_{1}}dx_{1}'\overline{\phi (x'_{1},x_{2}^{},\lambda )}
\phi (x_{1}',x_{2}^{},k),  \label{56}
\end{equation}
solves the equation
\begin{equation}
(i\partial _{x_{2}}^{}+\partial _{x_{1}}^{2}-\tilde{u}(x))\tilde{\phi}(x,k)=0
\label{Schrut}
\end{equation}
with transformed potential. It is natural to expect that this transformation
for complex parameter $\lambda $ would supply an example of a potential of
the type~(\ref{u+-}). Check of the fact that $\tilde{\phi}$ obeys
Eq.~(\ref{Schrut}) is based on the following identity for a pair of
arbitrary solutions $f(x)$ and $g(x)$ of the Eq.~(\ref{Schrut}):
\begin{equation}
i\partial _{x_{2}}^{}(\overline{f(x)}g(x))=-\partial _{x_{1}}^{}
W(\overline{f(x)},g(x)),  \label{1020}
\end{equation}
where Wronskian
\begin{equation}
W(\overline{f(x)},g(x))=\overline{f(x)}\partial
_{x_{1}}^{}g(x)-g(x)\partial _{x_{1}}^{}\overline{f(x)}  \label{39}
\end{equation}
was introduced.

In order to make equations~(\ref{53})--(\ref{56}) determined it is necessary
to substitute indefinite integrals by definite ones and to choose constants
of integration in a way that the potential $\tilde{u}$ is real and regular.
Thus in order to get transformations parametrized by constants and not by
functions of $x_{2}$ obeying some differential equations it is natural to
choose integrals in~(\ref{a55})--(\ref{56}) to be from infinity to $x_{1}$.
Then we can use the fact that the asymptotic behavior~(\ref{asymptx}) with
respect to $x_{1}$ is independent on $x_{2}$. On the other side, infinite
limits of these integrals require exact control of their convergency in the
recursive procedure. In addition one must write the solution $\tilde{\phi}$
as a linear combination of the Jost solution and of a solution corresponding
to discrete spectrum. The way to build the correct recursive procedures for
generating both solutions is suggested by the remark that
$\phi (x,\lambda )\Delta (x)^{-1}$ is solution of the Eq.~(\ref{Schrut}).

Thus we start with a regular rapidly decaying real potential $u(x)$ for
which all above mentioned elements of the direct and inverse problem are
given. In Sec. 2 we introduce an exact recursion procedure for an
arbitrary number of B\"{a}cklund transformations and corresponding Darboux
transformations for Jost solutions and solutions corresponding to the
discrete spectrum. We formulate conditions of reality and regularity of the
potentials constructed by these means and derive spectral data of the
transformed Jost solutions (in analogy with~\cite{darboux}, where
transformations of the continuous spectra were considered). In Sec. 3 this
recursion procedure is solved in terms of the original potential $u(x)$
and its Jost solution $\Phi(x,k)$. By these means we get a solution
depending on $(N+1)N$ complex parameters describing $N$ solitons
superimposed to a generic background. Necessary and sufficient conditions
satisfied by these parameters in order to get a regular and real solution
are explicitly given. In the case $u(x)\equiv 0$ we recover not only the
multisoliton solutions obtained in~\cite{KPsolitonsfirst}, but we are able
to identify all the regular and real solutions in the essentially more
general class of multisoliton solutions derived in~\cite{generalKPsolitons}.
This extension based on the condition that the values of the Jost solution
at the points of discrete spectrum are given as linear combinations of
values of this solution in the conjugated points essentially complicates the
whole construction, but the corresponding solutions are of the type
essential for applications (see~\cite{localized}--\cite{fractal}). In Sec. 4
we present the leading asymptotic behavior on the $x$-plane of the constructed
potentials. We show that this behavior is indeed of the type~(\ref{u+-}) and
that it essentially depends on the signs of the imaginary parts of parameters
$\lambda $ of the B\"{a}cklund transformations. This essentially distinguishes
the case of the two dimensional nonstationary Schr\"{o}dinger equation from
the case of the one dimensional stationary equation. In a forthcoming
publication these results will be applied to the study of perturbations of
such potentials, i.e.\ to the generic potentials of the type~(\ref{u+-}) by
means of the formulation of the scattering problem on nontrivial background
as suggested in~\cite{towards}.

\section{Recursion procedure}

Formulas~(\ref{53}),~(\ref{a55}), and~(\ref{56}) enable us to formulate the
recursion procedure for composing an arbitrary number of binary B\"{a}cklund
transformations. Indeed, if $\phi _{n}(x,k)$ solves equation~(\ref{Schru})
with potential $u_{n}(x)$,
\begin{equation}
i\partial _{x_{2}}^{}\phi _{n}^{}(x,k)+\partial _{x_{1}}^{2}\phi
_{n}^{}(x,k)=u_{n}^{}(x)\phi _{n}^{}(x,k),  \label{899}
\end{equation}
then we specify Eq.~(\ref{56}) for $\phi _{n+1}$ in the following way:
\begin{eqnarray}
&&\phi _{n+1}^{}(x,k)=\phi _{n}^{}(x,k)-\nonumber\\
&&\qquad -g_{n+1}^{}(x)\left[B_{n+1}'(k)+
\hspace{-32pt}\int\limits_{\qquad -(k_{\Im }+\lambda _{n+1\Im })
\infty }^{\,x_{1}} \hspace{-39pt}
dx_{1}'\overline{\phi _{n}^{}(x_{1}',x_{2}^{},\lambda
_{n+1}^{})}\phi _{n}^{}(x_{1}',x_{2}^{},k)\right] ,
\label{1000}
\end{eqnarray}
where we introduced notations
\begin{eqnarray}
&&g_{n+1}^{}(x)=\frac{\phi _{n}^{}(x,\lambda _{n+1}^{})}{\Delta
_{n+1}^{}(x)},  \label{1004} \\
&&\Delta _{n+1}^{}(x)=c_{n+1}^{}+ \hspace{-4pt}
\int\limits_{-\lambda _{n+1\Im }\infty }^{\,x_{1}} \hspace{-12pt}
dx_{1}'|\phi _{n}^{}(x_{1}',x_{2}^{},\lambda
_{n+1}^{})|_{}^{2},  \label{1005} \\
&&n=0,1,\ldots ,  \nonumber
\end{eqnarray}
and $c_{n+1}$ and $B_{n+1}'(k)$ are some $x$-independent constant
and function of $k$, correspondingly. Then $\phi _{n+1}$ has to be a
solution of the shifted ($n\rightarrow n+1$) Eq.~(\ref{899}) with potential
\begin{equation}
u_{n+1}^{}(x)=u_{n}^{}(x)-2\partial _{x_{1}}^{2}\log \Delta
_{n+1}^{}(x).  \label{1014}
\end{equation}
In what follows it is convenient to write all $\phi _{n}$ as sums of two
solutions of~(\ref{899}),
\begin{equation}
\phi _{n}^{}(x,k)=F_{n}^{}(x,k)+f_{n}^{}(x,k),  \label{1001}
\end{equation}
that are given by the recursion relations
\begin{eqnarray}
&&F_{n+1}^{}(x,k)=F_{n}^{}(x,k)-g_{n+1}^{}(x)\hspace{-32pt}
\int\limits_{\qquad -(k_{\Im }+\lambda _{n+1\Im })\infty }^{\,x_{1}}
\hspace{-39pt}
dx_{1}'\overline{\phi _{n}^{}(x_{1}',x_{2}^{},\lambda
_{n+1}^{})}F_{n}^{}(x_{1}',x_{2}^{},k),\qquad  \label{1003} \\
&&f_{n+1}^{}(x,k)=f_{n}^{}(x,k)-  \nonumber \\
&&\qquad -g_{n+1}^{}(x)\left[ B_{n+1}^{}(k)+\hspace{-4pt}
\int\limits_{-\lambda _{n+1\Im }\infty }^{\,x_{1}}\hspace{-10pt}
dx_{1}'\overline{\phi _{n}^{}(x_{1}',x_{2}^{},\lambda
_{n+1}^{})}f_{n}^{}(x_{1}',x_{2}^{},k)\right] ,  \label{10031} \\
&&n=0,1,\ldots ,  \nonumber
\end{eqnarray}
where functions $B_{n+1}^{}(k)$ differ from $B_{n+1}'(k)$ because
of the different bottom limits of integrals in Eqs.~(\ref{1003}) and~(\ref
{10031}). We also put
\begin{equation}
u_{0}^{}(x)=u(x),\qquad F_{0}^{}(x,k)=\Phi (x,k),\qquad f_{0}^{}(x,k)=0,
\label{1002}
\end{equation}
so that we start with the standard, real, regular, and rapidly decaying at
space infinity potential $u$.

Properties of all these objects are given in the following

\begin{Theorem}
\label{theorem1}Let the potential $u(x)$ in~(\ref{1}) be real and rapidly
decaying at space infinity and let $\Phi (x,k)$ be the Jost solution of this
equation, i.e.\ let $\Phi (x,k)$ be defined by~(\ref{3}) and~(\ref{4}). Let
also the sets of complex constants $\lambda _{1}$, $\lambda _{2}$, $\ldots
$, real nonzero constants $c_{1}$, $c_{2}$, $\ldots $ obey the following
conditions:
\begin{eqnarray}
&&\lambda _{n\Im }^{} \neq 0,\qquad n=1,2,\ldots ,  \label{1006} \\
&&|\lambda _{1\Im }^{}| >|\lambda _{2\Im }^{}|>\ldots ,  \label{1007} \\
&&\lambda _{n\Im }^{}c_{n}^{} >0,\qquad n=1,2,\ldots ,  \label{1008}
\end{eqnarray}
and let be given some functions $B_{1}(k)$, $B_{2}(k)$, $\ldots $
of the complex parameter $k$ .

\begin{enumerate}
\item  \label{t11}Integrals in~(\ref{1003}),~(\ref{10031}), and~(\ref{1005})
converge and define regular functions of $x$ and $k$, for $k_{\Im }^{}\neq
0,$ $k_{\Im }^{}+\lambda _{j\Im }\neq 0$ ($j=1,2,\ldots ,n$). Functions
$\Delta _{n+1}(x)$ have no zeroes on the $x$-plane.

\item  \label{t12}There exist nonzero limits
\begin{equation}
\lim_{x_{1}\rightarrow \pm k_{\Im }\infty
}e_{}^{ikx_{1}+ik^{2}x_{2}}F_{n}^{}(x,k)=A_{n}^{}(\pm ,k)  \label{1010}
\end{equation}
that are independent of $x_{2}$ and obey the recursion relation
\begin{eqnarray}
&&A_{0}(\pm ,k) =1,  \label{1011} \\
&&A_{n+1}^{}(\pm ,k) =\left[ 1+\theta (\pm k_{\Im }\lambda _{n+1\Im })
\frac{2i\lambda _{n+1\Im }^{}}{\overline{\lambda }_{n+1}^{}-k}\right]
A_{n}^{}(\pm ,k).\qquad \qquad  \label{1012}
\end{eqnarray}

\item  \label{t13}There exist finite nonzero limits
\begin{eqnarray}
&&\lim_{|x_{1}|\rightarrow \infty }e_{}^{i\lambda _{n+1\Re }x_{1}+|\lambda
_{n+1\Im }x_{1}|}g_{n+1}^{}(x)=  \nonumber \\
&&\qquad =\left\{
\begin{array}{ll}
\displaystyle\frac{2\lambda _{n+1\Im }^{}}{{}\overline{A_{n}^{}(+,
\lambda _{n+1}^{})}{}}e^{-i\bar{\lambda}_{n+1}^{2}x_{2}}, &
x_{1}^{}\rightarrow +\lambda _{n+1\Im }^{}\infty , \\
\displaystyle\frac{A_{n}^{}(-,\lambda _{n+1}^{})}{c_{n+1}^{}}
e^{-i\lambda _{n+1}^{2}x_{2}}, & x_{1}^{}\rightarrow -\lambda _{n+1\Im
}^{}\infty ,
\end{array}
\right.  \label{1013}
\end{eqnarray}

\item  \label{t14}There exist finite nonzero limits (for $n\geq 1$)
\begin{eqnarray}
&&\lim_{|x_{1}|\rightarrow \infty }e_{}^{i\lambda _{n\Re }x_{1}+|\lambda
_{n\Im }x_{1}|}f_{n}^{}(x,k)=  \nonumber \\
&&\qquad =-\left\{
\begin{array}{ll}
(B_{n}^{}(k)+b_{n}^{}(k))\displaystyle\frac{2\lambda _{n\Im }^{}}{
\overline{A_{n-1}^{}(+,\lambda _{n}^{})}}e^{-i\bar{\lambda}
_{n}^{2}x_{2}}, & x_{1}^{}\rightarrow +\lambda _{n\Im }^{}\infty , \\
B_{n}^{}(k)\displaystyle\frac{A_{n-1}^{}(-,\lambda _{n}^{})}{c_{n}^{}}
e^{-i\lambda _{n}^{2}x_{2}}, & x_{1}^{}\rightarrow -\lambda _{n\Im
}^{}\infty ,
\end{array}
\right.\qquad \label{10131}
\end{eqnarray}
where the functions
\begin{equation}
b_{n}^{}(k)=\mathop{\rm sgn}\nolimits\lambda _{n\Im } \hspace{-4pt}
\int\limits_{-\infty }^{\,+\infty }\hspace{-5pt}dx_{1}'
\overline{\phi _{n-1}^{}(x_{1}',x_{2}^{},\lambda _{n}^{})}
f_{n-1}^{}(x_{1}',x_{2}^{},k)  \label{10132}
\end{equation}
are $x_{2}$-independent.

\item  \label{t15}Functions $F_{n}(x,k)$, $g_{n}(x)$, $f_{n}(x,k)$, and
$\phi _{n}^{}(x,k)$ solve the nonstationary Schr\"{o}\-din\-ger
equation~(\ref{899}) with potential
\begin{equation}
u_{n}^{}(x)=u(x)-2\partial _{x_{1}}^{2}\log \prod_{j=1}^{n}\Delta
_{j}^{}(x).  \label{10133}
\end{equation}

\item  \label{t16}Functions
\begin{equation}
\Phi _{n}^{}(x,k)=\frac{F_{n}^{}(x,k)}{A_{n}^{}(-,k)},\qquad
n=1,2,\ldots ,  \label{1015}
\end{equation}
are the Jost solutions of the Eq.\ (\ref{899}) with potential~(\ref{1014}),
i.e.
\begin{equation}
\chi _{n}^{}(x,k)=e_{}^{ikx_{1}+ik^{2}x_{2}}\Phi _{n}^{}(x,k)
\label{1016}
\end{equation}
obey the modified integral equations~(\ref{12}) with $u_n(x)$
from~(\ref{10133}) substituted for $u(x)$.
\end{enumerate}
\end{Theorem}

The proof of the theorem is by induction on $n$. Therefore, it will be
sometime useful in referring to a formula (\#) depending on $n$ to make
explicit this dependence by writing (\#)$_{n}$ and then (\#)$_{n+1}$ when
the same formula is considered for $n\rightarrow n+1$.

We divide the proof into a sequence of Lemmas.

\begin{Lemma}
\label{Lemma1} Let for some $n\geq 1$ the functions $F_{n}$ and $f_{n}$ be
regular functions of $x$ and $k$, for $k_{\Im }^{}\neq 0$, \ $k_{\Im
}^{}+\lambda _{j\Im }\neq 0$ ($j=1,2,\ldots ,n$) and obey statements~\ref
{t12} and~\ref{t14} of the theorem, or let they be given by~(\ref{1002}) for
$n=0$. Let the functions $B_{n}(k)$ be regular functions of $k$ and $\lambda
_{n}$, $\lambda _{n+1}$, $c_{n+1}$ obey~(\ref{1006})--(\ref{1008}). Then for
$\phi _{n}$ defined in~(\ref{1001}) there exists the limit
\begin{equation}
\lim_{x_{1}\rightarrow \pm k_{\Im }\infty }e_{}^{ikx_{1}+ik^{2}x_{2}}\phi
_{n}^{}(x,k)=A_{n}^{}(\pm ,k),\qquad k\in \C,\qquad |k_{\Im}^{}|
<|\lambda _{n\Im }^{}|,  \label{1017}
\end{equation}
where $A_{n}$ is given in~(\ref{1010}) and $\Delta _{n+1}$ determined
by~(\ref{1005}) exists, has no zeroes on the $x$-plane and obeys the
asymptotic behavior
\begin{equation}
\Delta _{n+1}^{}(x)\rightarrow \left\{
\begin{array}{ll}
\displaystyle\frac{\left| A_{n}^{}(+,\lambda _{n+1}^{})\right| ^{2}}
{2\lambda _{n+1\Im }^{}}e_{}^{2\lambda _{n+1\Im
}^{}(x_{1}^{}+2\lambda _{n+1\Re }^{}x_{2}^{})}, & \qquad
x_{1}^{}\rightarrow +\lambda _{n+1\Im }^{}\infty , \\
c_{n+1}^{}, & \qquad x_{1}^{}\rightarrow -\lambda _{n+1\Im }^{}\infty ,
\end{array}
\right.  \label{1018}
\end{equation}
where in the case $n=0$ in agreement with~(\ref{1001}) $A_{0}=1$.
\end{Lemma}

{\sl Proof}. In the case $n=0$~(\ref{1017}) is nothing but the direct
consequence of~(\ref{4}) and~(\ref{1002}). If $n\geq 1$ then this asymptotic
behavior of $\phi _{n}$ trivially follows from~(\ref{1001}),~(\ref{1010})
and~(\ref{10131}). Thanks to this asymptotic behavior taking into account
that $|\lambda _{n\Im }|>|\lambda _{n+1\Im }|>0$ we get convergency of the
integral in~(\ref{1005}). Asymptotics~(\ref{1018}) follows
from~(\ref{1017}). Taking into account that $\Delta _{n+1}$ by definition is
a monotonous function of $x_{1}$ and that thanks to~(\ref{1008}) the signs
of both asymptotic limits in~(\ref{1018}) coincide, we see that this
function has no zeroes on the $x$-plane.

\begin{Lemma}
\label{Lemma2} Under conditions of Lemma~\ref{Lemma1} function $g_{n+1}$ as
defined by~(\ref{1004}) is regular and obeys the asymptotic properties~(\ref
{1013}).
\end{Lemma}

{\sl Proof} of this Lemma follows directly from results of Lemma~\ref{Lemma1}
as $\Delta _{n+1}$ has no zeroes and the asymptotic
behaviors~(\ref{1017}),~(\ref{1018}) guarantee that $g_{n+1}$ decays with
proper exponent for growing $x$. The exact values of the limits are also
readily obtained from~(\ref{1017}) and~(\ref{1018}).

\begin{Lemma}
\label{Lemma21}Under conditions of Lemma~\ref{Lemma1} the function $f_{n+1}$
as defined by~(\ref{10031}) is regular and obeys the asymptotic
properties~(\ref{10131})$_{n+1}$ and~(\ref{10132})$_{n+1}$.
\end{Lemma}

{\sl Proof}. Convergency of the integral in~(\ref{10031}) and regularity of
$f_{n+1}$ follow directly from Lemma~\ref{Lemma1} if we take into account
that by~(\ref{1007}) $|\lambda _{n+1\Im }|<|\lambda _{n\Im }|$. In fact,
thanks to this inequality, by~(\ref{1017})$_{n}$ and~(\ref{10131})$_{n}$ the
integrals $\int_{\pm \infty }^{x_{1}}\overline{\phi _{n}^{}(x'_{1},x_{2},
\lambda _{n+1})}f_{n}(x_{1}',x_{2},k)$ are convergent. Then
from~(\ref{10031}) we get
\begin{eqnarray*}
&&e_{}^{i\lambda _{n+1\Re }x_{1}+|\lambda _{n+1\Im
}x_{1}|}f_{n+1}^{}(x,k)=e_{}^{i\lambda _{n+1\Re }x_{1}+|\lambda _{n+1\Im
}x_{1}|}f_{n}^{}(x,k)- \\
&&\,-e_{}^{i\lambda _{n+1\Re }x_{1}+|\lambda _{n+1\Im
}x_{1}|}g_{n+1}^{}(x)\left[ B_{n+1}^{}(k)+ \hspace{-4pt}
\int\limits_{-\lambda _{n+1\Im }\infty }^{\,x_{1}}\hspace{-10pt}
dx_{1}'\overline{\phi _{n}^{}(x_{1}',x_{2}^{},\lambda
_{n+1}^{})}f_{n}^{}(x_{1}',x_{2}^{},k)\right] .
\end{eqnarray*}
The asymptotic property~(\ref{10131})$_{n}$ guarantees that the first term
in the r.h.s.\ goes to zero when $|x_{1}|\rightarrow \infty $
because of condition $|\lambda _{n+1\Im }|<|\lambda _{n\Im }|$. The first
factor of the second term has finite limits by Lemma~\ref{Lemma2} and thanks
to the mentioned convergency of the integral in brackets we prove~(\ref
{10131})$_{n+1}$ and~(\ref{10132})$_{n+1}$.

\begin{Lemma}
\label{Lemma3} Under conditions of Lemma~\ref{Lemma1} $F_{n+1}$ as defined
by~(\ref{1003}) is a regular function obeying asymptotic
property~(\ref{1010})$_{n+1}$ and $A_{n+1}$ is given by means of~(\ref{1012}).
\end{Lemma}

{\sl Proof}. Convergency of the integral in~(\ref{1003}) follows from the
asymptotic behaviors~(\ref{1010})$_{n}$ and~(\ref{1017})$_{n}$ and then
regularity of $F_{n+1}$ from Lemma~\ref{Lemma2}. In order to
prove~(\ref{1010})$_{n+1}$, first, we consider the asymptotic behavior of
the integral
term:
\begin{eqnarray}
&&e_{}^{i(k-\bar{\lambda}_{n+1})x_{1}+i(k^{2}-\bar{\lambda}_{n+1}^{2})x_{2}}
\hspace{-32pt}
\int\limits_{\qquad -(k_{\Im }+\lambda _{n+1\Im })\infty }^{\,x_{1}}
\hspace{-39pt} dx_{1}'\overline{\phi _{n}^{}(x_{1}',x_{2}^{},\lambda
_{n+1}^{})}F_{n}^{}(x_{1}',x_{2}^{},k)\rightarrow  \nonumber \\
&&\qquad \rightarrow \overline{A_{n}^{}(\pm \mathop{\rm sgn}
\nolimits(k_{\Im }^{}\lambda _{n+1\Im }^{}),\lambda _{n+1}^{})}
\frac{A_{n}^{}(\pm ,k)}{i(\overline{\lambda }_{n+1}^{}-k)},\qquad
x_{1}^{}\rightarrow \pm k_{\Im }^{}\infty ,  \label{1019}
\end{eqnarray}
where~(\ref{1010})$_{n}$ and~(\ref{1017})$_{n}$ were used. Let us now
write~(\ref{1003}) as
\begin{eqnarray*}
&&e_{}^{ikx_{1}+ik^{2}x_{2}}F_{n+1}^{}(x,k)=e_{}^{ikx_{1}+ik^{2}x_{2}}
F_{n}^{}(x,k)- \\
&&\qquad -e_{}^{\lambda _{n+1\Im }x_{1}-|\lambda _{n+1\Im }x_{1}|}\left(
e_{}^{i\lambda _{n+1\Re }x_{1}+|\lambda _{n+1\Im }x_{1}|+i\bar{\lambda}
_{n+1}^{2}x_{2}}g_{n+1}^{}(x)\right) \times \\
&&\qquad \times \left( e_{}^{i(k-\bar{\lambda}_{n+1})x_{1}+i(k^{2}-
\bar{\lambda}_{n+1}^{2})x_{2}} \hspace{-32pt}
\int\limits_{\qquad -(k_{\Im }+\lambda _{n+1\Im })\infty }^{\,x_{1}}
\hspace{-39pt}
dx_{1}'\overline{\phi _{n}^{}(x_{1}',x_{2}^{},\lambda
_{n+1}^{})}F_{n}^{}(x_{1}',x_{2}^{},k)\right).
\end{eqnarray*}
In the first term of the r.h.s. we can use~(\ref{1010})$_{n}$. The second
term is given as a product of three multipliers, each of them having finite
limit for $x_{1}$ going to infinity. Moreover, the first of them has nonzero
limit only for $x_{1}\rightarrow \lambda _{n+1\Im }\infty $.
Then~(\ref{1012})$_{n+1}$ follows from~(\ref{1013})$_{n}$ proved in
Lemma~\ref{Lemma2} and the asymptotic limit of the last multiplier given
above.

\begin{Lemma}
\label{Lemma4} If under the assumptions of Lemma~\ref{Lemma1} statement 5 of
Theorem~\ref{theorem1} is satisfied then this statement is also valid for
$n\rightarrow n+1$ with potential $u_{n+1}$ given in~(\ref{10133})$_{n+1}$.
\end{Lemma}

{\sl Proof}. By the inductive hypothesis $\phi _{n}$ obeys
\begin{equation}
(i\partial _{x_{2}}^{}+\partial _{x_{1}}^{2}-u_{n}^{}(x))\phi
_{n}^{}(x,k)=0  \label{1021}
\end{equation}
and the same equation is valid for $F_{n}$, $g_{n}$, and $f_{n}$. Thus by
identity~(\ref{1020})
\[
i\partial _{x_{2}}^{} \hspace{-32pt}
\int\limits_{\qquad -(k_{\Im }+\lambda _{n+1\Im })\infty }^{\,x_{1}}
\hspace{-39pt}
dx_{1}'\overline{\phi _{n}^{}(x_{1}',x_{2}^{},\lambda
_{n+1}^{})}F_{n}^{}(x_{1}',x_{2}^{},k)=-W(\overline{\phi
_{n}^{}(x,\lambda _{n+1}^{})},F_{n}^{}(x,k)),
\]
where the asymptotic behaviors~(\ref{1010}) and~(\ref{1017}) were taken into
account. Analogously from~(\ref{1005}) we get
\begin{equation}
i\partial _{x_{2}}^{}\Delta _{n+1}^{}(x)=-W(\overline{\phi
_{n}^{}(x,\lambda _{n+1}^{})},\phi _{n}^{}(x,\lambda _{n+1}^{})).
\label{10211}
\end{equation}
Then by~(\ref{1004}) and~(\ref{1021}) we get
\begin{equation}
(i\partial _{x_{2}}^{}+\partial
_{x_{1}}^{2}-u_{n+1}^{}(x))g_{n+1}^{}(x)=0  \label{1022}
\end{equation}
with potential $u_{n+1}^{}(x)$ defined as in~(\ref{1014}). Now, by~(\ref
{1001}),~(\ref{1003}) and~(\ref{10031}) it is easy to check that $F_{n+1}$,
$f_{n+1}$, and $\phi _{n+1}$ solve~(\ref{1022}).

\begin{Lemma}
\label{Lemma5} If under the assumptions of Lemma~\ref{Lemma1} statement~\ref
{t15} of Theorem~\ref{theorem1} is fulfilled then statement~\ref{t16} is
valid for $n\rightarrow n+1$.
\end{Lemma}

{\sl Proof}. We have to demonstrate that $\chi _{n+1}$ as defined
by~(\ref{1015}) and~(\ref{1016}) obeys integral equation~(\ref{12}) with
$u_{n+1}$ given in~(\ref{1014}). Since we know from Lemma~\ref{Lemma4} that
$F_{n+1}$ obeys differential equation~(\ref{1022}), we can write
\begin{eqnarray*}
&&\int dx'\partial _{x_{1}}^{}G_{0}^{}(x-x',k)u_{n+1}^{}(x')
\chi _{n+1}^{}(x',k)= \\
&&\qquad =\int dx'\partial _{x_{1}}^{}G_{0}^{}(x-x',k)(i\partial _{x_{2}'}^{}
+\partial _{x'_{1}}^{2}-2ik\partial _{x_{1}'}^{})\chi
_{n+1}^{}(x',k). \end{eqnarray*}
Taking into account that the $\partial _{x_{1}}$ derivative cancels the
slowly decaying terms in the asymptotic behavior of the Green's function we
can integrate by parts and then use~(\ref{61}). Thus
\begin{eqnarray*}
&&1+ \hspace{-2pt} \int\limits_{-k_{\Im }\infty }^{\qquad x_{1}}
\hspace{-6pt} dy_{1}^{}\int dx'\partial
_{y_{1}}G_{0}^{}(y_{1}^{}-x_{1}',x_{2}^{}-x'_{2},k)u_{n+1}^{}(x')
\chi _{n+1}^{}(x',k)= \\
&&\qquad =1+ \hspace{-2pt} \int\limits_{-k_{\Im }\infty }^{\qquad x_{1}}
\hspace{-6pt}
dy_{1}^{}\partial _{y_{1}}\chi _{n+1}^{}(y_{1}^{},x_{2}^{},k)= \\
&&\qquad =1+\chi _{n+1}^{}(x,k)-\lim_{x_{1}\rightarrow -k_{\Im }\infty}
\chi _{n+1}^{}(x,k).
\end{eqnarray*}
This proves the lemma thanks to~(\ref{1010}),~(\ref{1015}) and~(\ref{1016}).
In the same way it is easy to show that $g_{n}(x)$ and $f_{n}(x,k)$ obey the
corresponding homogeneous integral equation.

{\sl Proof of the Theorem~\ref{theorem1} }now follows from Lemmas~\ref
{Lemma1}--\ref{Lemma5} by induction on $n$ since thanks to~(\ref{1002}) the
theorem is valid for $n=0$.

\begin{Corollary}
\label{Corollary0}We have scalar products
\begin{eqnarray}
&&\int dx_{1}'\overline{F_{n}^{}(x_{1}',x_{2}^{},k+p)}
F_{n}^{}(x_{1}',x_{2}^{},\bar{k}) =2\pi \delta (p),\qquad p\in\R,
\label{10221} \\
&&\int dx_{1}'\overline{F_{n}^{}(x_{1}',x_{2}^{},k+p)}
f_{n}^{}(x_{1}',x_{2}^{},\bar{k}) =0.  \label{10222}
\end{eqnarray}
\end{Corollary}

{\sl Proof}. By means of~(\ref{1003}) we have
\begin{eqnarray*}
&&\overline{F_{n+1}^{}(x,k+p)}F_{n+1}^{}(x,\bar{k})=
\overline{F_{n}^{}(x,k+p)}F_{n}^{}(x,\bar{k})- \\
&&\qquad -\partial _{x_{1}}^{}\left[ \frac{1}{\Delta _{n+1}^{}(x)}
\hspace{-20pt}
\int\limits_{\qquad -(k_{\Im }+\lambda _{n+1\Im })\infty }^{\,x_{1}}
\hspace{-25pt} dx_{1}'\phi _{n}^{}(x_{1}',x_{2}^{},\lambda
_{n+1}^{})\overline{F_{n}^{}(x_{1}',x_{2}^{},k+p)}\times
\right.  \\
&&\qquad \left. \times \hspace{-20pt}
\int\limits_{\qquad (k_{\Im }-\lambda _{n+1\Im })\infty }^{\,x_{1}}
\hspace{-20pt} dx_{1}'\overline{\phi _{n}^{}(x_{1}',x_{2}^{},\lambda
_{n+1}^{})}F_{n}^{}(x_{1}',x_{2}^{},\bar{k})\right]
\end{eqnarray*}
It is easy to see that the integral of the last term is equal to zero and,
therefore, we get
\begin{equation}
\int dx_{1}'\overline{F_{n+1}^{}(x_{1}',x_{2}^{},k+p)}
F_{n+1}^{}(x_{1}',x_{2}^{},\bar{k})=\int dx_{1}'
\overline{F_{n}^{}(x_{1}',x_{2}^{},k+p)}F_{n}^{}(x'_{1},x_{2}^{},\bar{k})
\label{10223}
\end{equation}
that proves~(\ref{10221}) thanks to~(\ref{81}) and~(\ref{1002}). The second
equality is derived analogously.

\begin{Corollary}
\label{Corollary1} $F_{n}^{}(x,k)$ is an analytic function in the complex
plane of $k$ with possible discontinuity at the real axis and poles at points
$k=\overline{\lambda }_{j}$, \ $j=1,\ldots ,n$.
\end{Corollary}

{\sl Proof}. Indeed, by definition~(\ref{1003}) we see that $F_{n+1}$
inherits analyticity properties of $F_{n}$ and has an additional
discontinuity at $k_{\Im }=-\lambda _{n+1\Im }$. Thanks to~(\ref{1003})
and~(\ref{10221}),~(\ref{10222}) we have that
\begin{eqnarray}
&&F_{n+1}^{}(x,k_{\Re }^{}-i(\lambda _{n+1\Im }+0))-F_{n+1}^{}(x,k_{\Re}^{}
-i(\lambda _{n+1\Im }-0))=\nonumber\\
&&\qquad =2\pi \delta (k_{\Re }^{}-\lambda _{n+1\Re})
g_{n+1}^{}(x).  \label{1023}
\end{eqnarray}
Thus we see, that $F_{n+1}(x,k)$ has an additional pole at
$k=\bar{\lambda}_{n+1}$,
\begin{equation}
F_{n+1}^{}(x,k)=-\frac{ig_{n+1}^{}(x)}{k-\overline{\lambda }_{n+1}^{}}
+O(1),\qquad k\rightarrow \bar{\lambda}_{n+1}^{}.  \label{10231}
\end{equation}

\begin{Corollary}
\label{Corollary2} Because of~(\ref{1011}) and~(\ref{1012})
\begin{equation}
A_{n}(\pm ,k)=\prod_{j=1}^{n}\left( \frac{k-\lambda _{j}^{}}{k-
\overline{\lambda }_{j}^{}}\right) ^{\theta (\pm k_{\Im }\lambda _{j\Im })}.
\label{10232}
\end{equation}
Thus $A_{n}(-,k)$ is a meromorphic function discontinuous on the real axis
of $k$ that has simple poles at $k=\bar{\lambda}_{j}$, $j=1,\ldots ,n$.
\end{Corollary}

\begin{Corollary}
\label{Corollary3}Let us introduce the transmission coefficients
\begin{equation}
a_{n}(k)=\frac{A_{n}(+,k)}{A_{n}(-,k)}=\prod_{j=1}^{n}\left( \frac{k-\lambda
_{j}^{}}{k-\overline{\lambda }_{j}^{}}\right)
^{\sgn k_{\Im }\lambda _{j\Im }}.
\label{1028}
\end{equation}
Then like in the one-dimensional case
\begin{equation}
a_{n}^{}(k)=\lim_{x_1\rightarrow +k_{\Im }\infty }\chi _{n}^{}(x,k).
\label{1027}
\end{equation}
This function is analytic in the upper and bottom half planes with a
discontinuity at the real axis and has simple zeroes at points $k=\lambda
_{j}$ and $k=\overline{\lambda }_{j}$, $j=1,\ldots ,n$.
\end{Corollary}

\begin{Corollary}
\label{Corollary4}$\Phi _{n}^{}(x,k)$ is an analytic function in the
complex plane of $k$ with possible discontinuity at the real axis and
\begin{equation}
\Phi _{n+1}^{}(x,\bar{\lambda}_{n+1}^{})=-\frac{g_{n+1}^{}(x)}
{2\lambda _{n+1\Im }}\prod_{j=1}^{n}\left( \frac{\overline{\lambda }
_{n+1}^{}-\overline{\lambda }_{j}^{}}{\overline{\lambda }
_{n+1}^{}-\lambda _{j}^{}}\right) ^{\theta (\lambda _{n+1\Im }\lambda
_{j\Im })}.  \label{1029}
\end{equation}
\end{Corollary}

{\sl Proof} follows from Corollaries~\ref{Corollary1} and~\ref{Corollary2}
since the singularities of $F_{n}(x,k)$ in the complex plane are compensated
by normalization~(\ref{1015}). Then~(\ref{1029}) follows from~(\ref{10231}).

\begin{Corollary}
\label{Corollary5} Let the boundary values of $\Phi_{n}(x,k)$ and $a_{n}(k)$ at
the real axis be defined in analogy with~(\ref{sd1}). Then for these values we
have relation
\begin{equation}
\frac{\Phi _{n}^{+}(x,k)}{a_{n}^{+}(k)}=\Phi _{n}^{-}(x,k)+\int dp\,\Phi
_{n}^{-}(x,p){\cal F}_{n}^{}(p,k).  \label{1031}
\end{equation}
(cf.~(\ref{sd4})), where continuous part of the spectral data is given by
\begin{equation}
{\cal F}_{n}^{}(k,p)={\cal F}(k,p)\prod_{j=1}^{n}\left( \frac{(k-\lambda
_{j}^{})(p-\overline{\lambda }_{j}^{})}{(k-\overline{\lambda }_{j}^{})
(p-\lambda _{j}^{})}\right)_{}^{\theta (\lambda_{j\Im })},\qquad k,p\in \R.
\label{1030}
\end{equation}
\end{Corollary}

{\sl Proof}. For the boundary values of $F_{n}(x,k)$ at the real axis in analogy
with~(\ref{10223}) we derive
\begin{eqnarray}
&&\int dx_{1}'\overline{F_{n+1}^{\mp}(x'_{1},x_{2}^{},k)}
F_{n+1}^{\pm }(x_{1}',x_{2}^{},p) =\int dx_{1}' \overline{F_{n}^{\mp}
(x_{1}',x_{2}^{},k)}F_{n}^{\pm}(x_{1}',x_{2}^{},p)=\qquad\qquad   \nonumber \\
&&\qquad =2\pi \delta (k-p) \\
&&\int dx_{1}'\overline{F_{n+1}^{+}(x_{1}',x_{2}^{},k)}F_{n+1}^{+}
(x_{1}',x_{2}^{},p) =\int dx'_{1}
\overline{F_{n}^{+}(x_{1}',x_{2}^{},k)}F_{n}^{+}(x'_{1},x_{2}^{},p)=
\qquad \qquad\nonumber \\
&&\qquad =2\pi {\cal F}(k,p)+2\pi \delta (k-p),
\end{eqnarray}
where~(\ref{sd2}) and~(\ref{sd3}) were used. Then~(\ref{1031}) follows from
definition~(\ref{1015}) and~(\ref{10232}).

\section{Resolution of the recursion relations}

In order to resolve the recursion relations explicitly we introduce
\begin{eqnarray}
&&B_{l,m}^{}(x)= \hspace{-20pt}
\int\limits_{\qquad -(\lambda _{l\Im }+\lambda _{m\Im })\infty}^{\quad x_{1}}
\hspace{-20pt} dx_{1}'\overline{\Phi (x_{1}',x_{2}^{},\lambda _{l}^{})}\Phi
(x_{1}',x_{2}^{},\lambda _{m}^{}),  \label{609} \\
&&\beta _{l}^{}(x,k)= \hspace{-20pt}
\int\limits_{\qquad -(k_{\Im }+\lambda _{l\Im })\infty }^{\quad x_{1}}
\hspace{-20pt} dx_{1}'\overline{\Phi (x_{1}',x_{2}^{},\lambda _{l}^{})
}\Phi (x_{1}',x_{2}^{},k),  \label{671} \\
&&l,m=1,2,\ldots ,  \nonumber
\end{eqnarray}
so that
\begin{equation}
B_{l,m}^{}(x)=\beta _{l}^{}(x,\lambda _{m})=\overline{\beta
_{m}^{}(x,\lambda _{l})}.  \label{680}
\end{equation}
Let $B_{n}^{}(x)$ denotes the $n\times n$ matrix
\begin{equation}
B_{n}^{}(x)=\Vert B_{l,m}^{}(x)\Vert _{l,m=1,\ldots ,n}^{}  \label{658}
\end{equation}
and let us define the following row and two columns
\begin{eqnarray}
&&\Phi (x)=(\Phi (x,\lambda _{1}^{}),\ldots ,\Phi (x,\lambda _{n}^{})),
\label{678} \\
&&\beta (x,k)=(\beta _{1}^{}(x,k),\ldots ,\beta _{n}^{}(x,k))_{}^{\rm T},
\label{679} \\
&&\gamma (x,k)=(\gamma _{1}^{}(k),\ldots ,\gamma _{n}^{}(k))_{}^{\rm T},
\label{6791}
\end{eqnarray}
where subscript T means transposition and $\gamma _{n}(k)$ are some given
functions such that matrix
\begin{equation}
C_{n}^{}=\Vert c_{l,m}^{}\Vert _{l,m=1,\ldots ,n}^{},\qquad
c_{l,m}^{}=\gamma _{l}^{}(\lambda _{m}^{}),  \label{6792}
\end{equation}
is Hermitian. Let us denote
\begin{equation}
A_{n}^{}(x)=C_{n}^{}+B_{n}^{}(x),  \label{6811}
\end{equation}
that is also Hermitian matrix by construction.

In order to formulate conditions of regularity of the potential $u_n(x)$ we
introduce also matrices
\begin{equation}
C_{n}^{}(y)=\Vert \{c_{l,m}^{}|\;y\lambda _{l\Im }^{}>0,\;
y\lambda _{m\Im }^{}>0\}\Vert _{l,m=1,\ldots ,n}^{},
\label{6794}
\end{equation}
where $y$ is some real parameter, in fact its sign. Their sizes can be less
than $n\times n$, as they are constructed by removing from matrix $C_n$ those
rows and columns that do not obey the inequalities in~(\ref{6794}). These 
matrices are Hermitian also and if all rows and columns are removed, then we 
put by definition $\det C_{n}(y)=1$. Let also $\Lambda_{n}(y)$ denote 
matrix with entries $-i(\overline{\lambda}_{l}-\lambda _{m})^{-1}$ obeying 
the same properties as in~(\ref{6794}). One can show that
\begin{equation}
\det \Lambda_{n}^{}(y)=\prod_{l=1}^{n}(2\lambda _{l\Im }^{})^{-\theta
(y\lambda _{l\Im }^{})}\prod_{l,m=1 \atop l\neq m}^{n}{}\left|
\frac{\lambda _{l}^{}-\lambda _{m}^{}}{\overline{\lambda }_{l}^{}-\lambda
_{m}^{}}\right| _{}^{\theta (y\lambda _{l\Im }^{})\theta (y\lambda
_{m\Im }^{})}.  \label{6796}
\end{equation}

In these terms we impose the following conditions on the constants $c_{l,m}$:
\begin{equation}
\pm C_{n}^{}(\pm)>0  \label{67931}
\end{equation}
and prove the following

\begin{Theorem}
\label{theorem2}Let conditions~(\ref{1006}),~(\ref{1007}), and~(\ref{67931}) 
be fulfilled. Then for any $n=1,2,\ldots $ and $x$
\begin{equation}
\det A_{n}^{}(x)\prod_{l=1}^{n}\lambda _{l\Im }^{}>0,\qquad \det
A_{0}^{}(x)=1,  \label{67932}
\end{equation} and the solutions of the recursion
equations~(\ref{1014})--(\ref{1002}) are given by means of the following
relations:
\begin{eqnarray}
&&\Delta _{n}^{}(x)=\frac{\det A_{n}^{}(x)}{\det A_{n-1}^{}(x)},
\label{5623} \\
&&F_{n}^{}(x,k)=\frac{1}{\det A_{n}^{}(x)}\left|
\begin{array}{cc}
A_{n}^{}(x) & \beta (x,k) \\
\Phi (x) & \Phi (x,k)
\end{array}
\right| ,  \label{5624} \\
&&f_{n}^{}(x,k)=\frac{1}{\det A_{n}^{}(x)}\left|
\begin{array}{cc}
A_{n}^{}(x) & \gamma (k) \\
\Phi (x) & 0
\end{array}
\right| ,  \label{5625} \\
&&g_{n}(x)=\frac{-1}{\det A_{n}^{}(x)}\left|
\begin{array}{cc}
A_{n}^{}(x) & e_{n}^{} \\
\Phi (x) & 0
\end{array}
\right| ,\quad e_{n}^{}=(0,\ldots 0,1)_{}^{\rm T},  \label{56252} \\
&&u_{n}^{}(x)=u(x)-2\partial _{x_{1}}^{2}\log \det A_{n}^{}(x),
\label{56251} \\
&&n=1,2,\ldots   \nonumber
\end{eqnarray}
Moreover, for the constants $c_{n}$ in~(\ref{1005}) we have relation
\begin{equation}
c_{n+1}^{}=\frac{\det C_{n+1}^{}(\lambda^{}_{n+1\Im })}
{\det C_{n}^{}(\lambda^{}_{n+1\Im })},
\label{56253}
\end{equation}
where the matrices $C_{n}^{}(\pm)$ are defined in~(\ref{67931}).
\end{Theorem}

In order to prove that these relations resolve~(\ref{1014})--(\ref{1002})
we need to calculate all involved integrals. By~(\ref{1001}),~(\ref{5624})
and~(\ref{5625})
\begin{equation}
\phi _{n}^{}(x,k)=\frac{1}{\det A_{n}^{}(x)}\left|
\begin{array}{ll}
A_{n}^{}(x) & \gamma (k)+\beta (x,k) \\
\Phi (x) & \Phi (x,k)
\end{array}
\right| ,  \label{56256}
\end{equation}
so that
\begin{equation}
\phi _{n}^{}(x,\lambda _{n+1}^{})=\frac{1}{\det A_{n}^{}(x)}\left|
\begin{array}{ll}
A_{n}^{}(x) & A_{*,n+1}^{}(x) \\
\Phi (x) & \Phi (x,\lambda _{n+1}^{})
\end{array}
\right| ,  \label{56255}
\end{equation}
where we introduced the column
\begin{equation}
A_{*,n+1}^{}(x)=(A_{1,n+1}^{}(x),\ldots ,A_{n,n+1}^{}(x))_{}^{\rm T}
\label{562551}
\end{equation}
and used~(\ref{680}),~(\ref{6792}), and~(\ref{6811}) in order to write
$\gamma _{m}^{}(\lambda _{n+1})+\beta _{m}^{}(x,\lambda
_{n+1})=A_{m,n+1}(x)$. Let us also introduce the matrix
\begin{equation}
A_{n+1}^{}(x,k)=\left(
\begin{array}{ll}
A_{n}^{}(x) & \gamma (k)+\beta (x,k) \\
A_{n+1,*}(x) & \beta _{n+1}^{}(x,k)
\end{array}
\right) ,  \label{56257}
\end{equation}
where the row $A_{n+1,*}(x)$ is the transposition of the
column~(\ref{562551}). Below for any matrix $A$ we denote as $A^{(k,l)}$ the
same matrix with removed $l$-th row and $m$-th column, say,
\begin{equation}
A_{n+1}^{(l,m)}(x)=\Vert (A_{n+1}^{}(x,k))_{i,j}^{}
\Vert _{i,j=1,\ldots ,n, \atop i\neq l,\quad j\neq m}^{}.  \label{5617}
\end{equation}

\begin{Lemma}
\label{Lemma212}Let Theorem~\ref{theorem2} be valid for some $n$. Then
\begin{equation}\label{56171}
\overline{\phi _{n}^{}(x,\lambda _{n+1}^{})}\phi _{n}^{}(x,k)=\partial
_{x_{1}}^{}\frac{\det A_{n+1}^{}(x,k)}{\det A_{n}^{}(x)}.
\end{equation}
\end{Lemma}

{\sl Proof}. Thanks to notation~(\ref{56257}) we can write the expansions
of~(\ref{56255}) and~(\ref{56256}) with respect to the last row as
\begin{eqnarray}
&&\overline{\phi _{n}^{}(x,\lambda _{n+1}^{})}=\frac{1}{\det
A_{n}^{}(x)}\sum_{k=1}^{n+1}(-1)_{}^{n+1+k}\overline{\Phi (x,\lambda
_{k}^{})}\det A_{n+1}^{(k,n+1)}(x,k),  \label{5618} \\
&&\phi _{n}^{}(x,k)=\frac{1}{\det A_{n}^{}(x)}\biggl[
\sum_{l=1}^{n}(-1)_{}^{n+l+1}\Phi (x,\lambda _{l}^{})\det
A_{n+1}^{(n+1,l)}(x,k)+  \nonumber \\
&&\qquad +\Phi (x,k)\det A_{n+1}^{(n+1,n+1)}(x)\biggr].  \label{5619}
\end{eqnarray}
Then taking into account~(\ref{609}),~(\ref{680}),~(\ref{6811}),
and~(\ref{56257}) we get that
$\overline{\Phi (x,\lambda _{m}^{})}\Phi (x,\lambda_{l}^{})=
\partial _{x_{1}}^{}(A_{n+1}(x,k))_{m,l}$, $l\leq n$, and
$\overline{\Phi (x,\lambda _{m}^{})}\Phi (x,k)=\partial
_{x_{1}}^{}(A_{n+1}(x,k))_{m,n+1}$. Thus
\begin{eqnarray*}
&&\overline{\phi _{n}^{}(x,\lambda _{n+1}^{})}\phi _{n}^{}(x,k)= \\
&&\qquad =\sum_{k,l=1}^{n+1}(-1)_{}^{k+l}\frac{\left( \partial
_{x_{1}}^{}A_{n+1}^{}(x,k)\right) _{k,l}^{}}{\det_{}^{2}A_{n}^{}(x)}
\det A_{n+1}^{(k,n+1)}(x,k)\det A_{n+1}^{(n+1,l)}(x,k)
\end{eqnarray*}
and the statement of the lemma follows from the known property of
determinants: If $A_{n}$, $n=1,2,\ldots $, are $n\times n$ matrices
depending on some parameter and such that for every $n$ the matrix $A_{n}$ is
just the upper main minor of matrix $A_{n+1}$ then we have the identity
\begin{eqnarray}
&&(\det A_{n+1})'\det A_{n}^{}-(\det A_{n})'\det
A_{n+1}^{}=\nonumber\\
&&\qquad =\sum_{k,l=1}^{n+1}(-1)_{}^{k+l}A_{k,l}'\det A_{n+1}^{(k,n+1)}
\det A_{n+1}^{(n+1,l)},  \label{5621}
\end{eqnarray}
where prime denotes derivative with respect to this parameter.

\begin{Lemma}
\label{Lemma213}Let $A_{n}$ be defined in~(\ref{6811}). Then the leading
asymptotic behavior of its determinant for $|x_{1}|\rightarrow \infty $ and
$x_{2}$ fixed is given by
\begin{equation}
\det A_{n}(x)=\det C_{n}^{}(\mp )\det \Lambda_{n}^{}(\pm)
\prod_{j=1}^{n}\left| 1+e_{}^{-i\lambda _{j}x_{1}-i\lambda_{j}^{2}x_{2}}
\right| _{}^{2},\qquad x_{1}^{}\rightarrow \pm \infty ,
\label{5620}
\end{equation}
where matrices $C_{n}^{}(\mp )$ and $\det \Lambda_{n}^{}(\pm )$ are
defined in~(\ref{6794}) and~(\ref{6796}).
\end{Lemma}

{\sl Proof}. Let us introduce the $n\times n$ diagonal matrix
\begin{equation}
D_{n}^{}(x)={\rm diag}\left\{ 1+e_{}^{-i\lambda _{j}x_{1}-i\lambda
_{j}^{2}x_{2}}\right\} _{j=1}^{n}.  \label{56200}
\end{equation}
Then, taking into account that thanks to~(\ref{3}),~(\ref{asymptx}),
and~(\ref{671})
\begin{eqnarray}
&&e_{}^{i(k-\bar{\lambda}_{l})x_{1}+i(k^{2}-\bar{\lambda}_{l}^{2})x_{2}}
\beta _{l}^{}(x,k) \rightarrow \frac{1}{i(\overline{\lambda }_{l}^{}-k)}
\nonumber \\
&&e_{}^{i(\lambda _{m}-\bar{\lambda}_{l})x_{1}+i(\lambda _{m}^{2}-
\bar{\lambda}_{l}^{2})x_{2}}B_{l,m}^{}(x) \rightarrow \frac{1}
{i(\overline{\lambda }_{l}^{}-\lambda _{m}^{})}  \label{56202}
\end{eqnarray}
for $|x_{1}|\rightarrow \infty $ we derive that
\begin{equation}
\lim_{x_{1}\rightarrow \pm \infty }\overline{D_{n}^{}(x)}_{}^{-1}
A_{n}^{}(x)D_{n}^{}(x)_{}^{-1}=\alpha _{n}^{}(\pm ),
\label{56203}
\end{equation}
where the $n\times n$ matrices
$\alpha _{n}(\pm )=\Vert \alpha _{l,m}^{}(\pm)\Vert _{l,m=1,\ldots ,n}^{}$,
\ $\alpha _{0}=1$, are defined by means of their entries
\begin{equation}
\alpha _{l,m}^{}(\pm )=\theta (\lambda _{l\Im }^{}\lambda _{m\Im
}^{})\left( c_{l,m}^{}\theta (\mp \lambda _{l\Im }^{})+\frac{\theta
(\pm \lambda _{l\Im }^{})}{i(\overline{\lambda }_{l}^{}-\lambda
_{m}^{})}\right) .  \label{56201}
\end{equation}
We see that due to~(\ref{56201}) only entries ($l,m$) in the l.h.s.\ obeying
condition $\lambda _{l\Im }^{}\lambda _{m\Im}^{}>0$ can give nontrivial
limits. It is easy to notice that entries such that
$\lambda _{l\Im }^{}\lambda _{m\Im }^{}<0$ are decaying at least as
$e^{-|\lambda _{n\Im }x_{1}|}$, i.e.\ by~(\ref{1007}) as the lowest
exponential involved in the matrix $A_{n}$.

Taking into account the block structure of the matrix $\alpha_{n}(\pm)$
we get by~(\ref {67931})
\[
\det \alpha _{n}^{}(\pm )=\det C_{n}^{}(\mp )\det
\Lambda_{n}^{}(\pm ),
\]
that gives~(\ref{5620}).

\begin{Lemma}
\label{Lemma214} Let Theorem~\ref{theorem2} be valid for some $n$ and let
$\lambda _{1},\ldots ,\lambda _{n+1}$ obey~(\ref{1007}). Then for
$\Delta_{n+1}(x)$ defined in~(\ref{1005}) we have
equality~(\ref{5623})$_{n+1}$ where the coefficient $c_{n+1,n+1}$ of the
matrix $A_{n+1}(x)$ is given by
\begin{equation}
c_{n+1,n+1}^{}=c_{n+1}^{}+C_{n+1,*}^{}(\lambda^{}_{n+1\Im})C_{n}^{-1}
(\lambda^{}_{n+1\Im}) C_{*,n+1}^{}(\lambda^{}_{n+1\Im}),  \label{570}
\end{equation}
where $C_{n+1,*}(\lambda_{n+1\Im})$ and $C_{*,n+1}(\lambda_{n+1\Im})$ are the
last row and column of $C_{n+1}(\lambda_{n+1\Im})$. Moreover, matrix
$A_{n+1}(x)$ obeys property~(\ref{67932})$_{n+1}$.
\end{Lemma}

{\sl Proof}. In order to calculate the integral in~(\ref{1005}) we
use~(\ref{56255}), so $\gamma _{m}(\lambda _{n+1})=c_{m,n+1}$. Then by
Lemma~\ref{Lemma212}
\[
\overline{\phi _{n}^{}(x,\lambda _{n+1}^{})}\phi _{n}^{}(x,\lambda
_{n+1}^{})=\partial _{x_{1}}^{}\frac{\det A_{n+1}^{}(x)}{\det
A_{n}^{}(x)},
\]
where we used that in this case
$\det A_{n+1}(x,\lambda _{n+1})=\det A_{n+1}(x)-c_{n+1,n+1}\det A_{n}(x)$
by~(\ref{56257}). Then by Lemma~\ref{Lemma213} we see that the integral
in~(\ref{1005}) is convergent and that
\[
\Delta _{n+1}^{}(x)=c_{n+1}^{}+\frac{\det A_{n+1}^{}(x)}{\det
A_{n}^{}(x)}-\frac{\det C_{n+1}^{}(\lambda ^{}_{n+1\Im })\det
\Lambda_{n+1}^{}(-\lambda^{}_{n+1\Im })}{\det C_{n}^{}(\lambda^{}
_{n+1\Im })\det \Lambda_{n}^{}(-\lambda^{} _{n+1\Im })}.
\]
By definition~(\ref{6796})
\begin{equation}
\Lambda_{n+1}^{}(-\lambda^{}_{n+1\Im })=\Lambda_{n}^{}(-\lambda^{}_{n+1\Im }),
\label{5701}
\end{equation}
thus the determinants of $\Lambda$ cancel out and~(\ref{5623})$_{n+1}$ follows
by~(\ref{56253}). The latter one is equivalent to~(\ref{570}) thanks to the
following property of the determinants of bordered matrices:
\begin{equation}
\frac{1}{\det C_{n}^{}}\left|\begin{array}{ll}
C_{n}^{} & C_{*,n+1}^{}\\
C_{n+1,*}^{} & c_{n+1,n+1}^{}
\end{array}\right|=c_{n+1,n+1}^{}-C_{n+1,*}^{}C_{n}^{-1}C_{*,n+1}^{}.
\label{510}
\end{equation}
In order to prove~(\ref{67932})$_{n+1}$ let us mention first that
by~(\ref{6794})
\begin{equation}
C_{n+1}^{}(-\lambda^{} _{n+1\Im })=C_{n}^{}(-\lambda^{}_{n+1\Im }).
\label{57011}
\end{equation}
Second, let $v$ be an arbitrary $n$-column and let $v_{n+1}$ be an arbitrary
complex scalar. Then we have
\begin{equation} \label{57012}
\bigl(v^{\dag}_{},\overline{v}_{n+1}^{}\bigr)C^{}_{n+1}(\lambda^{}_{n+1\Im })
\left(\begin{array}{l} v\\ v_{n+1}^{}\end{array}\right)=
w^{\dag}_{}C^{}_{n}(\lambda^{}_{n+1\Im })w+|v_{n+1}^{}|_{}^{2}c^{}_{n+1},
\end{equation}
where in the last term~(\ref{56253}) was used and we denoted
\[
w=v+v_{n+1}^{}C^{-1}_{n}(\lambda^{}_{n+1\Im })C^{}_{*,n+1}
(\lambda^{}_{n+1\Im }).
\]
Thus we see that conditions~(\ref{67931})$_{n+1}$ for one of the signs 
trivially follow from~(\ref{57011}) and for the other sign are equivalent to 
the condition $c_{n+1}\lambda _{n+1\Im }>0$, that in its turn by 
Theorem~\ref{theorem1} is equivalent to the condition that 
$\Delta _{n+1}(x)$ has no zeroes on the $x$-plane and its sign is equal to 
the sign of $\lambda _{n+1\Im }$.  Then finally~(\ref{67932})$_{n+1}$ 
follows from~(\ref{5623})$_{n+1}$ and the proof of lemma is completed.

\begin{Lemma}
\label{Lemma215} Let Theorem~\ref{theorem2} be valid for some $n$ and let
$\lambda _{1},\ldots ,\lambda _{n+1}$ obey~(\ref{1007}). Then for $g_{n+1}(x)$
defined in~(\ref{1004}) we have equality~(\ref{56252})$_{n+1}$.
\end{Lemma}

{\sl Proof}. By~(\ref{1004}),~(\ref{5623}), and~(\ref{56255})
\begin{equation}
g_{n+1}^{}(x)=\frac{1}{\det A_{n+1}^{}(x)}\left|
\begin{array}{ll}
A_{n}^{}(x) & A_{*,n+1}^{}(x) \\
\Phi (x) & \Phi (x,\lambda _{n+1}^{})
\end{array}
\right|  \label{508}
\end{equation}
that is nothing but~(\ref{56252})$_{n+1}$ expanded with respect to the last
column.

\begin{Lemma}
\label{Lemma216} Let Theorem~\ref{theorem2} be valid for some $n$ and let
$\lambda _{1},\ldots ,\lambda _{n+1}$ obey~(\ref{1007}). Then for $F_{n+1}(x)$
as defined in~(\ref{1003}) we have equality~(\ref{5624})$_{n+1}$.
\end{Lemma}

{\sl Proof}. We need to calculate the integral with $F_{n}$ in~(\ref{1003}).
Thus in order to use~(\ref{56171}) for this lemma we have to choose 
in~(\ref{56256}) and, correspondingly, in~(\ref{56257}) all 
$\gamma_{l}(k)=0$. Then from Lemma~\ref{Lemma212} we get
\begin{equation}
F_{n+1}^{}(x,k)=F_{n}^{}(x,k)-g_{n+1}^{}(x) \hspace{-32pt}
\int\limits_{\qquad -(k_{\Im }+\lambda _{n+1\Im })\infty }^{\,x_{1}}
\hspace{-39pt} dx_{1}'\partial _{x_{1}'}^{}\frac{\det A_{n+1}^{}
(x_{1}',x_{2}^{},k)}{\det A_{n}^{}(x'_{1},x_{2}^{})}.  \label{509}
\end{equation}
Using the property~(\ref{510}) we get by~(\ref{56257}) and~(\ref{56200}) that
\begin{eqnarray*}
&&\frac{\det A_{n+1}^{}(x,k)}{\det A_{n}^{}(x)}=\beta _{n+1}^{}(x,k)-
\\
&&\qquad -\sum_{l,m=1}^{n}\frac{c_{n+1,l}^{}+B_{n+1,l}^{}(x)}
{1+e^{-i\lambda _{l}x_{1}-i\lambda _{l}^{2}x_{2}}}\frac{\beta _{m}^{}(x,k)}
{1+e^{i\bar{\lambda}_{m}x_{1}+i\bar{\lambda}_{m}^{2}x_{2}}}
\Biggl( \bigl(\overline{D}_{n}^{-1}(x)A_{n}^{}(x)D_{n}^{-1}(x)\bigr)_{}^{-1}
\Biggr)_{l,m}^{}.
\end{eqnarray*}
We need to consider the limit
$x_{1}\rightarrow -(k_{\Im }+\lambda _{n+1\Im})\infty $ of the r.h.s. Thanks
to the asymptotic behavior~(\ref{56202}) the first term goes to zero. By
Lemma~\ref{Lemma213} we know that the matrix
$(\overline{D}_{n}^{-1}A_{n}D_{n}^{-1})^{-1}$ has finite nonzero limit and
that entries of this matrix corresponding to
$\lambda _{l\Im }\lambda_{m\Im}<0$, as it was mentioned for the inverse
matrix in the proof of Lemma~\ref{Lemma213}, exponentially decay as
$e^{-|\lambda _{n\Im }x_{1}|}$.
Then we can write for the asymptotics of $\beta _{m}(x,k)$ that
$e^{k_{\Im}x_{1}}=e^{(k_{\Im }+\lambda _{n+1\Im })x_{1}}e^{-\lambda_{n+1\Im}
x_{1}}$ where the first factor decays by the sign of infinity and  because
of condition~(\ref{1007}) the second factor is majorized by some of the
decaying exponents.

Now by means of~(\ref{5624})$_n$ and~(\ref{508}) we get from~(\ref{509})
\begin{equation}
F_{n+1}^{}(x,k)=\frac{1}{\det A_{n}^{}(x)\det A_{n+1}^{}(x)}\left|
\begin{array}{cc}
A_{n}^{}(x) & Z(x,k) \\
\Phi (x) & z_{n+1}^{}(x,k)
\end{array}
\right| ,  \label{511}
\end{equation}
where we introduced the column $Z=(z_{1},\ldots ,z_{n})^{\rm T}$ and the
entry $z_{n+1}$ as follows:
\begin{eqnarray*}
&&Z =\beta (x,k)\det A_{n+1}^{}(x)-A_{*,n+1}^{}(x)\left|
\begin{array}{cc}
A_{n}^{}(x) & \beta (x,k) \\
A_{n+1,*}^{}(x) & \beta _{n+1}^{}(x,k)
\end{array}
\right| , \\
&&z_{n+1}^{} =\Phi (x,k)\det A_{n+1}^{}(x)-\Phi (x,\lambda
_{n+1}^{})\left|
\begin{array}{cc}
A_{n}^{}(x) & \beta (x,k) \\
A_{n+1,*}^{}(x) & \beta _{n+1}^{}(x,k)
\end{array}
\right| .
\end{eqnarray*}
By means of~(\ref{56257}) with $\gamma (k)=0$ this can be rewritten as
\begin{eqnarray*}
&&z_{j}^{}=(A_{n+2}^{}(x,k))_{j,n+2}^{}\det A_{n+2}^{(n+2,n+2)}(x,k)- \\
&&\qquad -(A_{n+2}^{}(x,k))_{j,n+1}^{}\det A_{n+2}^{(n+2,n+1)}(x,k) \\
&&z_{n+1}^{}=\Phi (x,k)\det A_{n+2}^{(n+2,n+2)}(x,k)-\Phi (x,\lambda
_{n+1}^{})\det A_{n+2}^{(n+2,n+1)}(x,k).
\end{eqnarray*}
We see that the determinant in~(\ref{511}) is unchanged if all
$z_{1},\ldots ,z_{n+1}$ are replaced with
\begin{eqnarray*}
&&\tilde{z}_{j}^{}= \sum_{l=1}^{n+2}(-1)_{}^{n+l}(A_{n+2}^{}(x,k))_{j,l}^{}
\det A_{n+2}^{(n+2,l)}(x,k),\qquad k=1,\ldots ,n, \\
&&\tilde{z}_{n+1}^{}=\Phi (x,k)\det A_{n+2}^{(n+2,n+2)}(x,k)+
\sum_{l=1}^{n+1}(-1)_{}^{n+l}\Phi (x,\lambda_{l}^{})\det A_{n+2}^{(n+2,l)}
(x,k),
\end{eqnarray*}
as all additional terms are just columns proportional to some other columns
of determinant in~(\ref{511}). On the other side we see that all
$\tilde{z}_{j}$ for $j=1,\ldots ,n$ are equal to determinants of the matrix
$A_{n+2}(x,k)$ with row $n+2$ replaced with $l$-th row of the same matrix.
So all of them are equal to zero and thus the determinant in~(\ref{511}) is
equal to $\tilde{z}_{n+1}\det A_{n}(x)$. On the other side $\tilde{z}_{n+1}$
is nothing but expansion of the determinant in~(\ref{5624})$_{n+1}$ with
respect to the last row. The lemma is proved.

\begin{Lemma}
\label{Lemma217} Let Theorem~\ref{theorem2} be valid for some $n$ and let
$\lambda _{1},\ldots ,\lambda _{n+1}$ obey~(\ref{1007}). Then for $f_{n+1}(x)$
as defined in~(\ref{10031}) we have equality~(\ref{5625})$_{n+1}$.
\end{Lemma}

{\sl Proof}. We need to calculate the integral with $f_{n}$ in~(\ref{10031}).
Thus we have to replace in~(\ref{56256}) in the last column $\beta (x,k)$
and $\Phi (x,k)$ with zeros. Correspondingly, Lemma~\ref{Lemma212} must be
used with a matrix $A_{n+1}(x,k)$ as in~(\ref{56257}) with only
$\gamma$-terms in the last column and zero on the bottom place. Using the
same consideration as in Lemma~\ref{Lemma216} we prove that for such matrix
$A_{n+1}(x,k)$ the limit of $\det A_{n+1}(x,k)\det A_{n}^{-1}(x)$ for
$x_{1}\rightarrow -\lambda _{n+1\Im }\infty $ is finite, and then the
integral in~(\ref{10031}) is convergent. Continuing in this way we
prove~(\ref{5625})$_{n+1}$ and get relation between $B_{n+1}(k)$ and $\gamma
_{n+1}(k)$. We omit these details, as in what follows solutions $f_{n}(x,k)$
are not used.

{\sl Proof of the Theorem~\ref{theorem2}} follows by induction on the base
of the lemmas proved above, if we notice that for $n=1$ the formulation of
the theorem coincides with formulas~(\ref{1004})--(\ref{10031}) for $n=0$ if 
we take into account~(\ref{1002}) and condition~(\ref{1008}) for $n=1$. As 
well Eq.~(\ref{56251})$_{n+1}$ follows from~(\ref{1014}) 
and~(\ref{56251})$_{n}$ thanks to~(\ref{5623})$_{n+1}$, i.e.\ thanks to
Lemma~\ref{Lemma214}. Let us emphasize that reality and regularity of
potentials $u_{n} $ for all $n$ are equivalent to conditions that matrices
$C_{n}$ are Hermitian and obey property~(\ref{67931}). These properties are 
independent on the original potential $u(x)$, so they coincide with the 
conditions given in~\cite{generalKPsolitons}, where nondiagonal matrix $C_n$ 
was introduced first time for the case $u(x)\equiv 0$.

\begin{Corollary}\label{Corollary21}
\begin{equation}
\Phi _{n}^{}(x,\lambda _{m}^{})=\sum_{l=1}^{n}d_{l,m}^{}\Phi_{n}^{}
(x,\overline{\lambda }_{l}^{}),  \label{512}
\end{equation}
where we introduced the constants
\begin{equation}
d_{l,m}^{}=2c_{l,m}^{}\lambda _{l\Im }^{}\prod_{j=1 \atop j\neq l}^{n}
\left( \frac{\overline{\lambda }_{l}^{}-\lambda _{j}^{}}
{\overline{\lambda }_{l}^{}-\overline{\lambda }_{j}^{}}\right) ^{\theta
(\lambda _{l\Im }\lambda _{j\Im })}\prod_{j=1}^{n}\left( \frac{\lambda
_{m}^{}-\overline{\lambda }_{j}^{}}{\lambda _{m}^{}-\lambda _{j}^{}}\right)
^{\theta (-\lambda _{m\Im }\lambda _{j\Im })}.  \label{513}
\end{equation}
\end{Corollary}

{\sl Proof}. Taking~(\ref{81}) into account we use~(\ref{671})
and~(\ref{679}) in~(\ref{5624}) in order to write
\[
F_{n}^{}(x,k_{\Re }^{}-i\lambda _{l\Im }^{}+i0)-F_{n}^{}(x,k_{\Re}^{}-
i\lambda _{l\Im }^{}-i0)=-2\pi \delta (k_{\Re }^{}-\lambda _{l\Re}^{})
\varphi _{l}^{}(x),\qquad l=1,\ldots ,n,
\]
where in analogy with~(\ref{56252}) we introduced
\begin{equation}
\varphi _{l}^{}(x)=\frac{-1}{\det A_{n}^{}(x)}\left|
\begin{array}{cc}
A_{n}^{}(x) & e_{l}^{} \\
\Phi (x) & 0
\end{array}
\right| ,  \label{514}
\end{equation}
where $e_{l}^{}=(0,\ldots 0,1,0,\ldots 0)_{}^{\rm T}$ is a column with
1 on the $l$-th place only. Thus as we already know from Corollary~\ref
{Corollary1} $F_{n}$ has poles at points $k=\overline{\lambda }_{l}$ and
\[
\mathop{\rm res}_{k=\overline{\lambda }_{l}}F_{n}^{}(x,k)=-i\varphi
_{l}^{}(x).
\]
Then by~(\ref{1015}) and Corollary~\ref{Corollary2}
\[
\Phi _{n}^{}(x,\overline{\lambda }_{l}^{})=\frac{\varphi _{l}^{}(x)}
{2\lambda _{l\Im }^{}}\prod_{j=1 \atop j\neq l}^{n}\left(
\frac{\overline{\lambda }_{l}^{}-\overline{\lambda }_{j}^{}}
{\overline{\lambda}_{l}^{}-\lambda _{j}^{}}\right) ^{\theta (\lambda _{l\Im }
\lambda _{j\Im })}.
\]
On the other side directly by~(\ref{5624}), again taking into
account~(\ref{1015}) and Corollary~\ref{Corollary2} we have
\[
\Phi _{n}^{}(x,\lambda _{m}^{})=\sum_{l=1}^{n}c_{l,m}^{}\varphi
_{l}^{}(x)\prod_{j=1}^{n}\left( \frac{\lambda _{m}^{}-\overline{\lambda }
_{j}^{}}{\lambda _{m}^{}-\lambda _{j}^{}}\right) ^{\theta (-\lambda
_{m\Im }\lambda _{j\Im })}
\]
that proves the statement.

This corollary completes the formulation of the inverse problem for the Jost
solution $\Phi _{n}$ as its discontinuity at the real axis was given by
Corollary~\ref{Corollary5}.

\begin{Corollary}\label{Corollary22} Condition~(\ref{1007}) for the final
formulas~(\ref{5624}) and~(\ref{56251}) can be omitted.
\end{Corollary}

{\sl Proof}. Indeed, this condition was relevant only for the proof that these
formulas obey recursion procedure of the Theorem~\ref{theorem1}. On the other
side Eqs.~(\ref{5624}),~(\ref{56251}), and~(\ref{67931}) are invariant under
any permutation of $\lambda_j$'s obeying~(\ref{1006}).

\section{Properties of potentials.}

We proved that potentials given by~(\ref{56251}) are real and regular and
these properties are equivalent to Hermiticity of the matrix $C_{n}$ defined
in~(\ref{6792}) and conditions~(\ref{67931}). Here we demonstrate that these
potentials are of the type~(\ref{u+-}). So we have to study their asymptotic
behavior when $x_{1}$ is replaced with $x_{1}-2\mu x_{2}$ and
$x_{2}\rightarrow \infty $, where $x_{1}$ is fixed and $\mu $ is a parameter
determining the direction of asymptotics on the $x$-plane. In the same way
like in Lemma~\ref{Lemma213} it is easy to prove that the leading term of the
potential is decaying for all
$\mu \neq \lambda _{1\Re },\ldots ,\lambda _{n\Re }$. Taking into account
that the original potential is rapidly decaying we have by~(\ref{56251}) for
this limit
\begin{equation}
\lim_{x_{2}\rightarrow \infty }u_{n}(x_{1}^{}-2\lambda _{j\Re
}^{}x_{2}^{},x_{2}^{})=-2\lim_{x_{2}\rightarrow \infty }\partial
_{x_{1}}^{2}\log \det A_{n}^{}(x_{1}^{}-2\lambda _{j\Re
}^{}x_{2}^{},x_{2}^{}).  \label{610}
\end{equation}
In analogy with Lemma~\ref{Lemma213} we introduce the diagonal matrix
\[
\Gamma _{l,m}^{}=\delta _{l,m}^{}\delta _{l,j}^{}+\delta
_{l,m}^{}(1-\delta _{l,j}^{})\left( 1+e^{-i\lambda _{l}x_{1}-i(\lambda
_{l}-\lambda _{j\Re })^{2}x_{2}-i\lambda _{j\Im }^{2}x_{2}}\right) .
\]
It is obvious that in~(\ref{610}) we can replace matrix $A_{n}$ with
matrix $\overline{\Gamma }^{-1}A_{n}\Gamma ^{-1}$ without changing the value
of the limit. In what follows we denote the transformations of the matrix
$A_{n}$ that do not affect the limit~(\ref{610}) by sign $\simeq $. Let now
introduce in analogy with~(\ref{56201}) the $(n-1)\times (n-1)$ matrix
\begin{eqnarray}
&&\alpha (j,\pm )=\Vert \alpha _{l,m}^{}(j,\pm )\Vert _{l,m=1,\ldots
,n,\;l,m\neq j}^{},  \label{611} \\
&&\alpha _{l,m}^{}(j,\pm )=c_{l,m}^{}\theta (\mp \lambda _{l\Im
}^{}(\lambda _{l\Re }^{}-\lambda _{j\Re }^{}))\theta (\mp \lambda
_{m\Im }^{}(\lambda _{m\Re }^{}-\lambda _{j\Re }^{}))+  \nonumber \\
&&\qquad +\frac{\theta (\pm \lambda _{l\Im }^{}(\lambda _{l\Re
}^{}-\lambda _{j\Re }^{}))\theta (\pm \lambda _{m\Im }^{}(\lambda
_{m\Re }^{}-\lambda _{j\Re }^{}))}{i(\overline{\lambda } _{l}^{}-
\lambda _{m}^{})}.  \label{612}
\end{eqnarray}
This matrix has block structure and in analogy
with~(\ref{6794})--(\ref{6796}) we introduce
\begin{eqnarray}
&&C(j,x_{2})=\Vert \{c_{l,m}^{}|\;\lambda _{l\Im }^{}(\lambda _{l\Re
}^{}-\lambda _{j\Re }^{})x_{2}^{}>0,\;\lambda _{m\Im }^{}(\lambda
_{m\Re }^{}-\lambda _{j\Re }^{})x_{2}^{}>0\}\Vert ,  \label{613} \\
&&\Lambda(j,x_{2})=\Vert \{-i(\overline{\lambda }_{l}^{}-\lambda
_{m}^{})_{}^{-1}|\;\lambda _{l\Im }^{}(\lambda _{l\Re }^{}-\lambda
_{j\Re }^{})x_{2}^{}>0,  \nonumber \\
&&\qquad \lambda _{m\Im }^{}(\lambda _{m\Re }^{}-\lambda _{j\Re }^{})
x_{2}^{} >0\}\Vert ,  \label{614} \\
&&l,m=1,\ldots ,n,\qquad l,m\neq j,  \nonumber
\end{eqnarray}
and again we put the determinant of a matrix that has no entries equal to 1.
Then we get
\[
\det \alpha (j,\pm )=\det C(j,\mp )\det \Lambda(j,\pm ).
\]
For $\Lambda(j,\pm )$ we have explicitly
\begin{eqnarray}
&&\det \Lambda(j,\pm )=\prod_{l=1 \atop l\neq j}^{n}(2\lambda _{l\Im}^{})^{-\theta
(\pm \lambda _{l\Im }(\lambda _{l\Re }-\lambda _{j\Re}))}\times   \nonumber \\
&&\qquad \times \prod_{l,m=1 \atop l\neq m,\, l,m\neq j}^{n}\left|
\frac{\lambda _{l}^{}-\lambda _{m}^{}}{\overline{\lambda }
_{l}^{}-\lambda _{m}^{}}\right| _{}^{\theta (\pm \lambda _{l\Im
}(\lambda _{l\Re }-\lambda _{j\Re }))\theta (\pm \lambda _{m\Im }(\lambda
_{m\Re }-\lambda _{j\Re }))}  \label{617}
\end{eqnarray}
and thanks to the condition~(\ref{67931}) of regularity
$\det \alpha (j,\pm)\neq 0$. Dividing
$\det (\overline{\Gamma }^{-1}A_{n}\Gamma ^{-1})$ by
$\det \alpha $ and using the analog of Eq.~(\ref{510}) we get thanks to the
asymptotic behavior~(\ref{56202}) that
\begin{eqnarray*}
&&\lim_{x_{2}\rightarrow \pm \infty }\det A_{n}^{}(x_{1}^{}-2\lambda
_{j\Re }^{}x_{2}^{},x_{2}^{})\simeq c_{j,j}^{}+\frac{e_{}^{2\lambda _{j\Im }
x_{1}}}{2\lambda _{j\Im }^{}}- \\
&&-\sum_{l,m=1 \atop l,m\neq j}^{n}\left[ c_{j,l}^{}\theta (\mp
\lambda _{l\Im }^{}(\lambda _{l\Re }^{}-\lambda _{j\Re }^{}))+
\frac{e_{}^{i\bar{\lambda}_{j}x_{1}}}{i(\overline{\lambda }_{j}^{}-\lambda
_{l}^{})}\theta (\pm \lambda _{l\Im }^{}(\lambda _{l\Re }^{}-\lambda
_{j\Re }^{}))\right] {}\times  \\
&&\times (\alpha (j,\pm )_{}^{-1})_{l,m}^{}\left[ c_{m,j}^{}\theta
(\mp \lambda _{m\Im }^{}(\lambda _{m\Re }^{}-\lambda _{j\Re }^{}))+
\frac{e_{}^{-i\lambda _{j}x_{1}}}{i(\overline{\lambda }_{m}^{}-\lambda
_{j}^{})}\theta (\pm \lambda _{m\Im }^{}(\lambda _{m\Re }^{}-\lambda
_{j\Re }^{}))\right] {}.
\end{eqnarray*}
Thanks to~(\ref{611}) and~(\ref{612}) the elements of the matrix
$(\alpha^{-1})_{l,m}$ are proportional to corresponding $\theta $-functions.
Then taking~(\ref{613}) and~(\ref{614}) into account we can write
\begin{eqnarray*}
&&\lim_{x_{2}\rightarrow \pm \infty }\det A_{n}^{}(x_{1}^{}-2\lambda
_{j\Re }^{}x_{2}^{},x_{2}^{})\simeq c_{j,j}^{}- \\
&&\qquad -\sum_{l,m=1 \atop l,m\neq j}^{n}c_{j,l}^{}\theta (\mp
\lambda _{l\Im }^{}(\lambda _{l\Re }^{}-\lambda _{j\Re }^{}))(C(j,\mp)_{}^{-1})
_{l,m}^{}c_{m,j}^{}\theta (\mp \lambda _{m\Im }^{}(\lambda
_{m\Re }^{}-\lambda _{j\Re }^{}))+ \\
&&\qquad +e_{}^{2\lambda _{j\Im }x_{1}}\Biggl[ \frac{1}{2\lambda _{j\Im}^{}}-
\sum_{l,m=1 \atop l,m\neq j}^{n}\frac{\theta (\pm \lambda
_{l\Im }^{}(\lambda _{l\Re }^{}-\lambda _{j\Re }^{}))}
{i(\overline{\lambda }_{j}^{}-\lambda _{l}^{})}(\Lambda(j,\pm)_{}^{-1})
_{l,m}^{} \times  \\
&&\qquad \times \frac{\theta (\pm \lambda _{m\Im }^{}(\lambda _{m\Re
}^{}-\lambda _{j\Re }^{}))}{i(\overline{\lambda }_{m}^{}-\lambda
_{j}^{})}\Biggr] .
\end{eqnarray*}
Let us introduce now the matrix $\widehat{C}(j,\pm )$ constructed by
removing from the matrix $C_{n}=\Vert c_{l,m}\Vert $ all rows
and columns with numbers $l\neq j$ such that
$\pm \lambda _{l\Im }(\lambda _{l\Re }-\lambda _{j\Re })<0$. Let the matrix
$\widehat{\Lambda}(j,\pm )$ be the same as $\widehat{C}(j,\pm )$ with $c_{l,m}$
substituted by $-i(\overline{\lambda}_{l}-\lambda _{m})^{-1}$. Then again
by~(\ref{510}) we get
\[
\det A_{n}^{}(x_{1}^{}-2\lambda _{j\Re
}^{}x_{2}^{},x_{2}^{})\simeq \frac{\det \widehat{C}(j,\mp )}{\det
C(j,\mp )}+e_{}^{2\lambda _{j\Im }x_{1}}\frac{\det \widehat{\Lambda}(j,\pm )}
{\det \Lambda(j,\pm )}.
\]
By~(\ref{617})
\begin{equation}
\frac{\det \widehat{\Lambda}(j,\pm )}{\det \Lambda(j,\pm )}=\frac{1}{2\lambda
_{j\Im}^{}}\prod_{l=1 \atop l\neq j}^{n}\left| \frac{\lambda
_{l}^{}-\lambda _{j}^{}}{\overline{\lambda }_{l}^{}-\lambda _{j}^{}}
\right| _{}^{2\theta (\pm \lambda _{l\Im }(\lambda _{l\Re }-\lambda _{j\Re}))},
\label{618}
\end{equation}
and using~(\ref{610}) we have finally that
\begin{equation}
\lim_{x_{2}\rightarrow \pm \infty }u_{n}(x_{1}^{}-2\lambda _{j\Re}^{}x_{2}^{},
x_{2}^{})=-\frac{2\lambda _{j\Im }^{2}}{\cosh^2_{} (\lambda
_{j\Im }^{}x_{1}^{}+\varepsilon _{j,\pm }^{})},  \label{619}
\end{equation}
where
\[
e_{}^{2\varepsilon _{j,\pm }}=\frac{\det C(j,\mp )}{\det \widehat{C}(j,\mp)}
\frac{\det \widehat{\Lambda}(j,\pm )}{\det \Lambda(j,\pm )}.
\]
This proves that the potentials constructed by means of the binary
B\"{a}cklund transformations indeed give nontrivial examples of the
class~(\ref{u+-}) since they are not decaying along directions
$x_1+2\lambda_{j\Re}x_2=$const.  The two rays belonging to each direction
are mutually shifted by
\begin{eqnarray}
&&e_{}^{2(\varepsilon _{j,+}-\varepsilon _{j,-})}=\frac{\det \widehat{C}
(j,+)\det C(j,-)}{\det \widehat{C}(j,-)\det C(j,+)}\times\nonumber\\
&&\qquad \times\prod_{l=1 \atop l\neq j}^{n}
\left| \frac{|\lambda _{j\Re }^{}-\lambda _{l\Re}^{}|
-i(\lambda _{j\Im }^{} \sgn (\lambda _{l\Re }-\lambda _{j\Re })-
|\lambda _{l\Im }^{}|)}{|\lambda
_{j\Re }^{}-\lambda _{l\Re }^{}|-i(\lambda _{j\Im }^{} \sgn
(\lambda _{l\Re }-\lambda _{j\Re })+|\lambda _{l\Im }^{}|)}\right|^{2}_{} ,
\label{620}
\end{eqnarray}
where we used~(\ref{618}).

All consideration here was done under assumption that all $\lambda_{j}$'s are
in a generic situation, in particular, that their real parts are all
different. In the situation when, say, $\lambda_{j\Re}=\lambda_{m\Re}$ the
limit in~(\ref{610}) does not exist (for a generic matrix $C_n$) as at large
$x_2$ there are oscillating terms. So in this case the potential does not
belong to the class~(\ref{u+-}).

Here only the leading asymptotic behavior of the potentials
$u_n(x)$ were considered. More detailed investigation performed
in~\cite{towards} for the case $n=1$ shows that even in that simple
situation the asymptotic behavior gets some rational corrections in the
higher terms. Moreover, these corrections depend on the sign of
$\lambda_{1\Im}$ if the original potential $u(x)$ is a nontrivial one. This
dependence on signs of the imaginary parts of $\lambda_{j}$'s is also
obvious here. Indeed, formulation of the inverse problem in~(\ref{1031})
involves spectral data ${\cal F}_n(k,p)$ that include only $\lambda_{j}$'s
with positive imaginary parts. Such dependence of the potential on these
signs is a specific two dimensional feature and as well as nondiagonal
matrix in~(\ref{512}) it has no analog in the one dimensional case. We plan to
discuss these aspects in more detail in a forthcoming publication.

\end{document}